\documentclass[epj]{webofc}
\usepackage[utf8]{inputenc}
\usepackage[varg]{txfonts}   
\usepackage{booktabs}
\usepackage{xcolor}
\definecolor{darkred}{rgb}{0.4,0.0,0.0}
\definecolor{darkgreen}{rgb}{0.0,0.4,0.0}
\definecolor{darkblue}{rgb}{0.0,0.0,0.4}
\usepackage[bookmarks,linktocpage,colorlinks,
    linkcolor = darkred,
    urlcolor  = darkblue,
    citecolor = darkgreen]{hyperref}
\usepackage{subfigure}
\usepackage{physics}
\usepackage{siunitx}

\wocname{EPJ Web of Conferences}
\woctitle{Lattice2017}
\newcommand{\be}{\begin{equation}}
\newcommand{\ee}{\end{equation}}
\newcommand{\bea}{\begin{eqnarray}}
\newcommand{\eea}{\end{eqnarray}}

\newcommand{\fm}{\mathrm{fm}}

\newcommand{\Eq}[1]{Eq.~(\ref{#1})}

\newcommand{\Sect}[1]{Section~\ref{#1}}

\begin{document}
%
\selectlanguage{english}
\title{%
Multi-boson block factorization of fermions
}
\author{%
  \firstname{Leonardo} \lastname{Giusti}\inst{1,2,3}\fnsep\thanks{Speaker, \email{Leonardo.Giusti@cern.ch}. Preprint numbers: CERN-TH-2017-195, DESY 17-170, HIM-2017-07
  }
\and
\firstname{Marco} \lastname{C\`e}\inst{4,5,6} \and
\firstname{Stefan}  \lastname{Schaefer}\inst{7}}
\institute{%
Theoretical Physics Department, CERN, Geneva, Switzerland
\and
Dipartimento di Fisica, Universit\`a di Milano-Bicocca, Piazza della Scienza 3, 
I-20126 Milano, Italy
\and
INFN, Sezione di Milano-Bicocca, Piazza della Scienza 3, 
I-20126 Milano, Italy
\and
Helmholtz-Institut Mainz,
Johannes Gutenberg-Universit\"at, Staudingerweg 18, D-55128 Mainz, Germany
\and
Scuola Normale Superiore, Piazza dei Cavalieri 7, I-56126 Pisa, Italy
\and
INFN, Sezione di Pisa, Largo B. Pontecorvo 3, I-56127 Pisa, Italy
\and
John von Neumann Institute for Computing (NIC), DESY, Platanenallee 6, D-15738
Zeuthen, Germany
}
\abstract{%
The numerical computations of many quantities of theoretical and phenomenological
interest are plagued by statistical errors which increase exponentially with the
distance of the sources in the relevant correlators.
Notable examples are baryon masses and matrix elements, the hadronic vacuum
polarization and the light-by-light scattering contributions to the muon $g-2$,
and the form factors of semileptonic $B$ decays. Reliable and precise determinations
of these quantities are very difficult if not impractical with state-of-the-art
standard Monte Carlo integration schemes.
I will review a recent proposal for factorizing the fermion determinant
in lattice QCD that leads to a local action in the gauge field and in
the auxiliary boson fields. Once combined with the corresponding factorization
of the quark propagator, it paves the way for multi-level Monte Carlo integration
in the presence of fermions opening new perspectives in lattice QCD. Exploratory
results on the impact on the above mentioned observables will be presented.
}
\maketitle
\section{Introduction}\label{intro}
Over the last three decades we have had an extraordinary conceptual, algorithmic
and technical progress in numerical lattice gauge theory which have led to the
simulation of Quantum Chromodynamics (QCD) with quark masses at the
physical point, see Ref.~\cite{Luscher:2010ae} for a recent review. Lattice
QCD became a theoretical femtoscope for studying the dynamics of the strong
interactions in Nature. It opened the window on quantities not accessible to
experiments which may help understanding the underlying dynamical mechanisms
of the theory. The interesting chiral regime of QCD became accessible to
non-perturbative computations.

The femtoscope, however, is still rather crude. Often we compute what we can
and not what we would like to. With state of the art techniques, numerical
computations of hadronic correlation functions suffer from signal-to-noise ratios
which decrease exponentially with the time separation of the sources, notable
exceptions being the propagators of non-singlet pseudoscalar mesons. For connected
Wick contractions, the problem can be traced back to the fact that, on a typical
gauge configuration, the quark propagator decreases approximatively as
$\exp{-M_\pi |y-x|/2}$ at asymptotically large distances $|y-x|$, while the
expectation value of a generic hadron correlator decays much
faster~\cite{Parisi:1983ae,Lepage:1989hd}.
This problem afflicts many computations at the forefront of research in lattice QCD:
the hadronic vacuum polarization and light-by-light
scattering contributions to the muon $g-2$, the amplitudes of leptonic and semileptonic
$B$ decays, masses and matrix elements of (multi) baryons states, etc. It is
timely to solve this problem so to be able to extract the maximum information from
the new experimental results expected in the coming years.

The conceptual framework for a solution has already been introduced in
bosonic theories. The multi-level Monte Carlo integration takes advantage of the fact that,
when the action and the observables depend locally on the integration variables,
the degradation of the signal-to-noise ratio with the distance of the sources can
be avoided by measuring independently the local building blocks of the observables.
This leads to an impressive acceleration of the
simulations~\cite{Albanese:1987ds,Luscher:2001up,Meyer:2002cd,DellaMorte:2007zz,
DellaMorte:2008jd,DellaMorte:2010yp}, and fully solves the problem in some cases.

It is not straightforward, however, to formulate multi-level algorithms for systems
with fermions. Once they have been analytically integrated out in the path
integral, the manifest locality of the action and of the observables is lost.
The fermion determinant and propagator are non-local functionals of the background
gauge field.
The aim of this talk is to review a recently proposed factorization of the fermion
determinant in lattice QCD that leads to a bosonic theory with a local
action in the block gauge, pseudofermion and multi-boson fields~\cite{Ce:2016ajy}. Together with
the factorization of the fermion observables presented in Ref.~\cite{Ce:2016idq},
this opens the way for multi-level simulations of QCD.  Exploratory
results on the impact on the above mentioned computations will also be reviewed.

\section{Signal/noise ratio in lattice QCD}
At large time distances  $|y_0-x_0|$, the zero-momentum propagator of a
non-singlet pseudoscalar meson and its
variance decay as
\be
C_\pi(y_0,x_0) =  \langle W_\pi(y_0,x) \rangle \propto e^{-M_\pi|y_0-x_0|}\, , 
 \qquad \quad \sigma^2_\pi (y_0,x_0) \propto e^{-2 M_\pi|y_0-x_0|}\, ,  
\ee
where
\be
W_\pi(y_0,x) = \sum_{\vec y} \Tr\left\{Q^{-1}(y,x) [Q^{-1}(y,x)]^\dagger \right\}\; ,
\ee
and the Hermitian-Dirac operator is defined as\footnote{For definitiveness, in
these proceedings we will only consider the case of $D$ being the massive Wilson-Dirac
operator with or without $O(a)$-improvement term.} $Q=\gamma_5 D$. This is so because
the mean and the width of the distribution of the positive stochastic
variable $\Tr\left\{Q^{-1}(y,x) [Q^{-1}(y,x)]^\dagger \right\}$ decay exponentially with
the same exponent at large distances, which suggests that
{\it configuration by configuration} in the representative ensemble
it holds
\be\label{eq:confbyconf}
\Tr\left\{Q^{-1}(y,x) [Q^{-1}(y,x)]^\dagger \right\}\propto
e^{-M_\pi|y-x|}\; .
\ee
This is confirmed by numerical results on the lattice.
As a consequence, the typical size of a connected Wick contraction at large time
distances is $\exp{-n M_\pi |y_0-x_0|/2}$, with $n$ being the number of
quark propagators, while the expectation value of a generic hadron correlator
decays much faster because hadron masses are naturally  much larger than
$M_\pi$~\cite{Parisi:1983ae,Lepage:1989hd}.  For disconnected contractions, the problem
is even worse due to the vacuum contribution to the variance. As we will see in the next sections,
the cause of the problem, i.e. Eq.~(\ref{eq:confbyconf}), is also a key ingredient
of its solution.

Today the exponential degradation of the signal-to-noise ratio sets the limits of
many computations of theoretical and phenomenological interest. In the remaining
part of this section we list some examples which at present are the object of an
intense theoretical and experimental research activity.

\subsection{Baryon correlators}
The nucleon two-point function at zero momentum $C_N$ is the prototype example
of this sort. The signal-to-noise ratio squared decreases as 
\be\label{eq:Nuclexp}
\frac{C^2_N(y_0,x_0)}{\sigma^2_N(y_0,x_0)} \propto e^{-(2 M_N - 3 M_\pi)|y_0-x_0|}\; ,
\ee
where $|y_0-x_0|$ is the time-distance of the sources and $(2 M_N- 3 M_\pi)$ is as big as
$7.4$~fm$^{-1}$ at the physical point. The number of configurations needed to reach a given
statistical precision thus increases with that exponential factor.
Analogous considerations hold for three-point (and higher)
baryonic correlation functions. For a precise and accurate determination of $g_A$ at the
physical point, for instance, the chiral effective theory suggests that a time separation of $2.0-2.5$~fm
is needed between the axial vector current and the nucleon interpolating
operators~\cite{Bar:2015zwa,Bar:2016uoj,Bar:2017gqh}. At present, typical time separations affordable
in numerical calculations are, instead, between $1.0$ and $1.5$~fm. The problem becomes
more and more severe for correlation functions of fields with higher and higher baryon number.

\begin{figure}[t!]
  \centering
  \includegraphics[width=0.45\columnwidth]{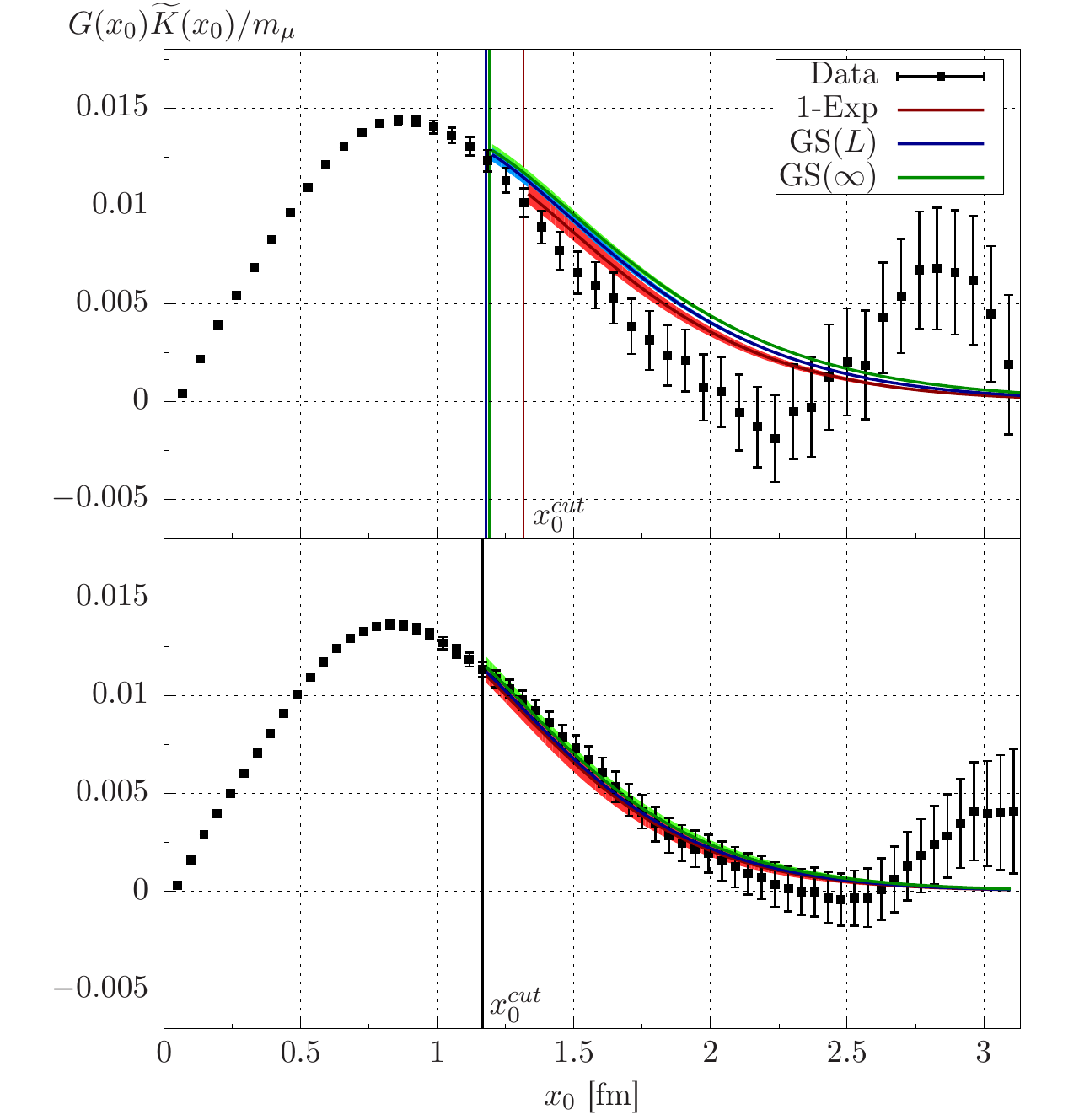}
  \caption{Data for the light quark contribution to the integrand
    $\widetilde K(x_0; m_\mu)\, G^{ud}(x_0)$ in QCD with two light dynamical flavours, scaled
    in units of the muon mass for two lattices with pion masses of approximatively $190$ (top) and
    $270$~MeV (bottom). The coloured bands, which show the various methods to constrain the long-distance
    behaviour, start at the respective value of $x_0^{\rm cut}$ as indicated by the vertical lines. By courtesy
    of Ref.~\cite{DellaMorte:2017dyu}.
  }
\label{fig:g-2N}
\end{figure}

\subsection{Vector correlators}
For the non-singlet vector two-point function at zero momentum $C_\rho$, the
signal-to-noise ratio squared goes as 
\be
\frac{C^2_\rho(y_0,x_0)}{\sigma^2_\rho(y_0,x_0)} \propto  e^{-2 (M_\rho - M_\pi)|y_0-x_0|}\; ,
\ee
where $m_\rho$ is the lightest asymptotic state in that channel. In the singlet case,
the exponential degradation is again worse due to the vacuum contribution to the variance.
This fact prevents a precise determination of, among other quantities, the Hadronic vacuum polarization (HVP)
and the Hadronic light-by-light (HLbL) contributions to the muon $g-2$ on the lattice
at the physical point. The HVP can indeed be written as~\cite{Bernecker:2011gh}
\be\label{eq:ahvp}
a_\mu^{\rm HVP} = \left(\frac{\alpha}{\pi}\right)^2 \int_0^{\infty} d x_0\, \widetilde K(x_0; m_\mu)\, G(x_0)\; , 
\ee
where $\alpha$ is the electromagnetic coupling constant, $\widetilde K(x_0; m_\mu)$ is a known
analytic function, and $G(x_0)$ is the correlation function of two electromagnetic
currents at a temporal distance $x_0$, see Ref.~\cite{DellaMorte:2017dyu} for unexplained
notation. The non-singlet contribution to $a_\mu^{\rm HVP}$ of the $u$ and $d$ quarks is shown in
Fig.~\ref{fig:g-2N} for pion masses of approximatively $190$ (top) and $270$~MeV (bottom)
respectively\footnote{These results are chosen among many others to illustrate the problem, see
Ref.~\cite{Lehener:lattice2017} for a detailed discussion on $a_\mu^{\rm HVP}$ at this
conference.}~\cite{DellaMorte:2017dyu}. In those plots the effect of the
exponential decrease of the signal-to-noise ratio with the distance $x_0$ of the sources
is evident. As a result, the contribution to the integral in Eq.~(\ref{eq:ahvp}) is computed
from data up to $x_0^{\rm cut}=1.1$-$1.4$~fm only,
while the rest is estimated by the Gounaris-Sakurai based extension
of the vector correlator~\cite{DellaMorte:2017dyu}. The final statistical and dominant systematic errors
turn out to be approximatively $5$ and $2.5$ percent respectively. If the signal could be kept well
under control up to time distances of $2.5$-$3.0$~fm or more, then one would be able to reach
the percent precision or better within QCD.

\begin{figure}[t!]
  \centering
  \includegraphics[width=0.4\columnwidth]{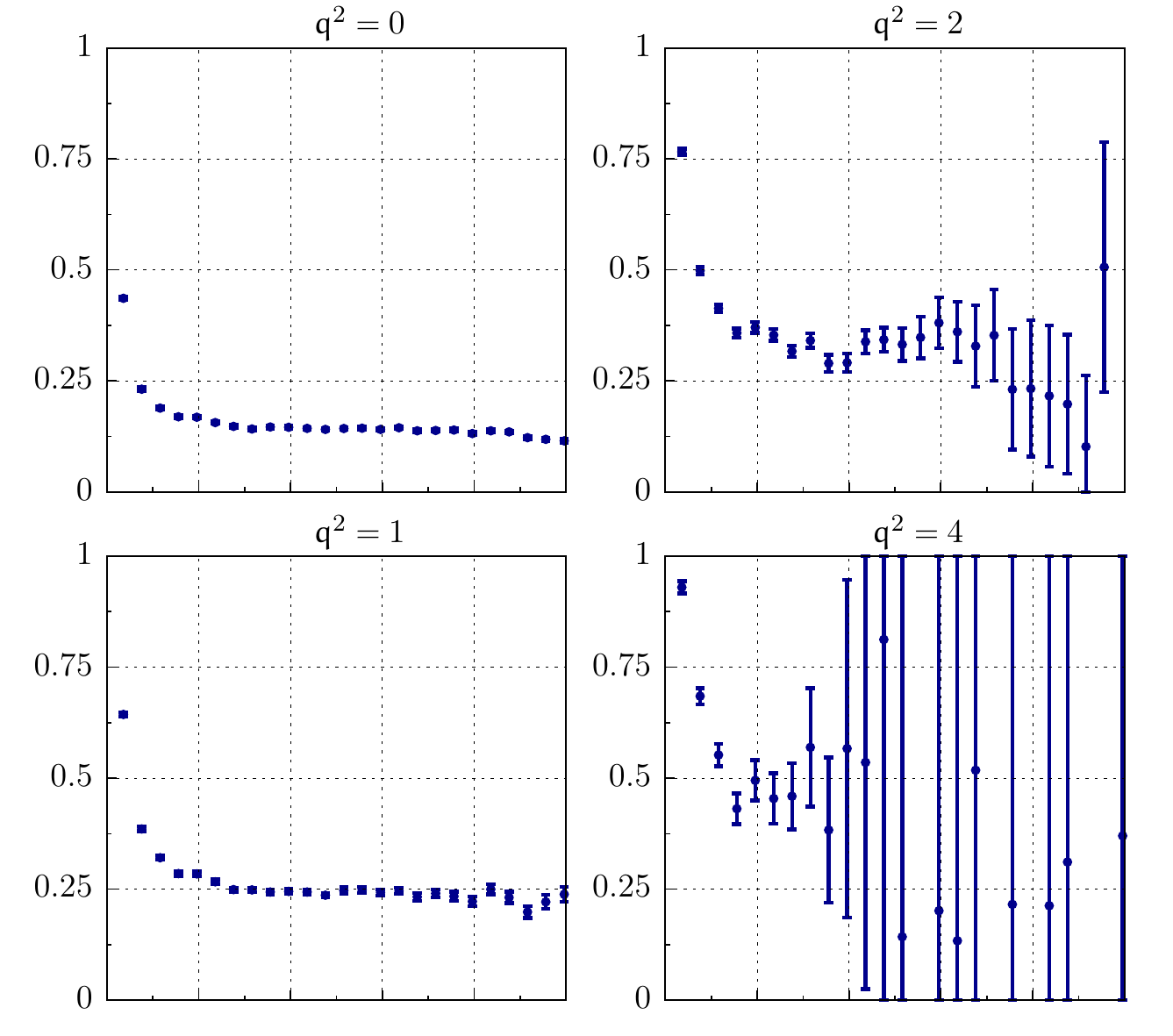}
  \includegraphics[width=0.4\columnwidth]{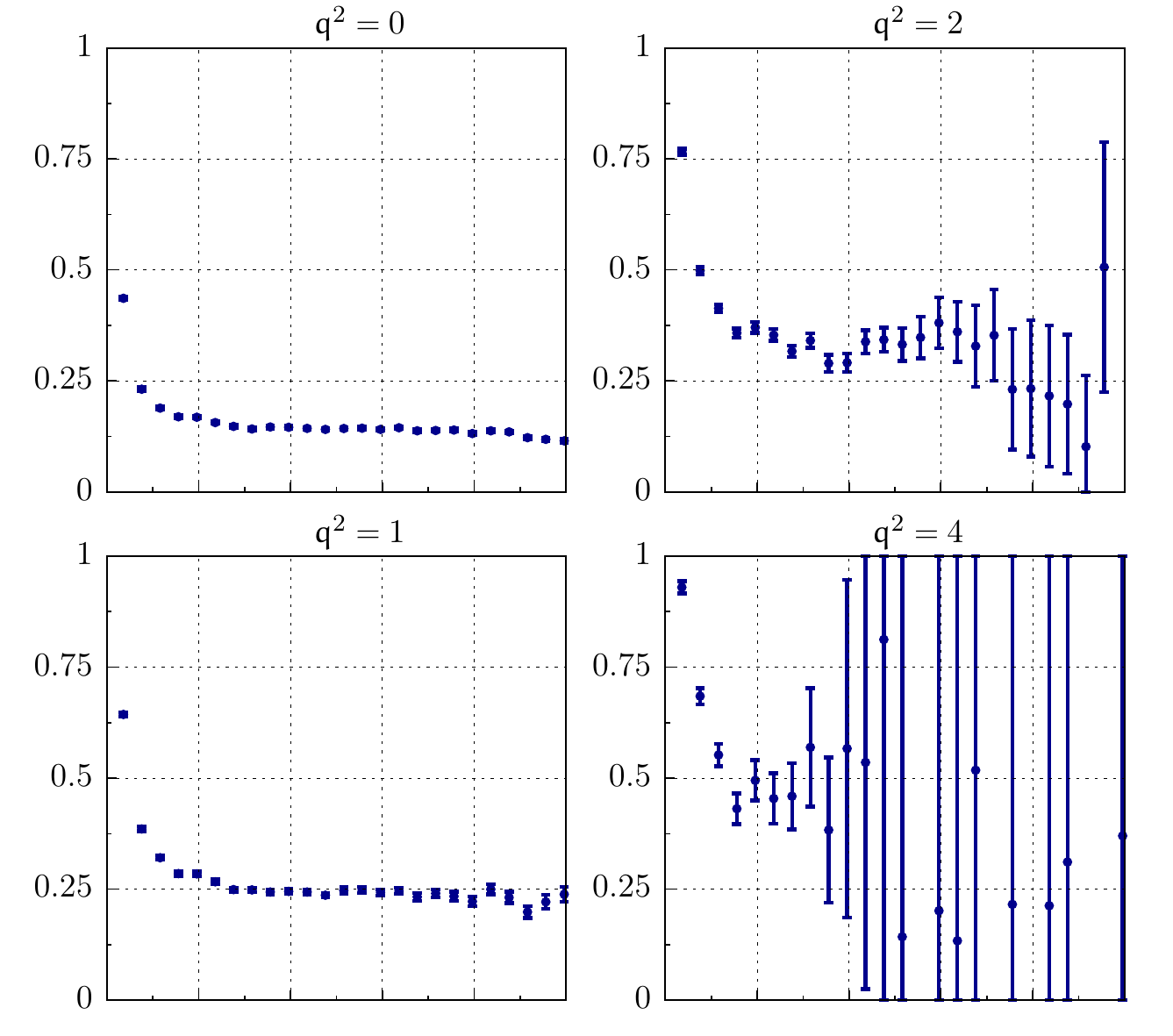}
  \caption{Pion effective energies for $|p|\approx 840$~MeV (left) and $|p|\approx 1200$~MeV (right)
           as a function of the time distance of the sources in units of
           the lattice spacing. By courtesy of Ref.~\cite{DellaMorte:2012xc}.}
\label{fig:moti24}
\end{figure}

\subsection{Non-zero momentum correlators}
Non-zero momentum correlators suffer from the exponential degradation of
the signal-to-noise ratio as well. For the pseudoscalar mesons, the signal-to-noise ratio squared
goes as 
\be
\frac{C^2_{\pi,{\vec p}}(y_0,x_0)}{\sigma^2_{\pi,{\vec p}}(y_0,x_0)} \propto  e^{-2 (E_\pi(\vec p) - M_\pi)|y_0-x_0|}\; . 
\ee
An example of effective energies corresponding to $|p|\approx 840$ (left) and $|p|\approx 1200$ MeV (right)
from Ref.~\cite{DellaMorte:2012xc} are shown in Fig.~\ref{fig:moti24}. The exponential decrease
of the signal-to-noise ratio is evident in these data. The same would happen if the momentum
is given to the mesons by imposing twisted boundary conditions in the spatial directions. Needless to
say this is one of the basic building blocks entering the Wick contractions of hadronic and
semileptonic decays, see below.

\begin{figure}[t!]
  \centering
  \includegraphics[width=0.49\columnwidth]{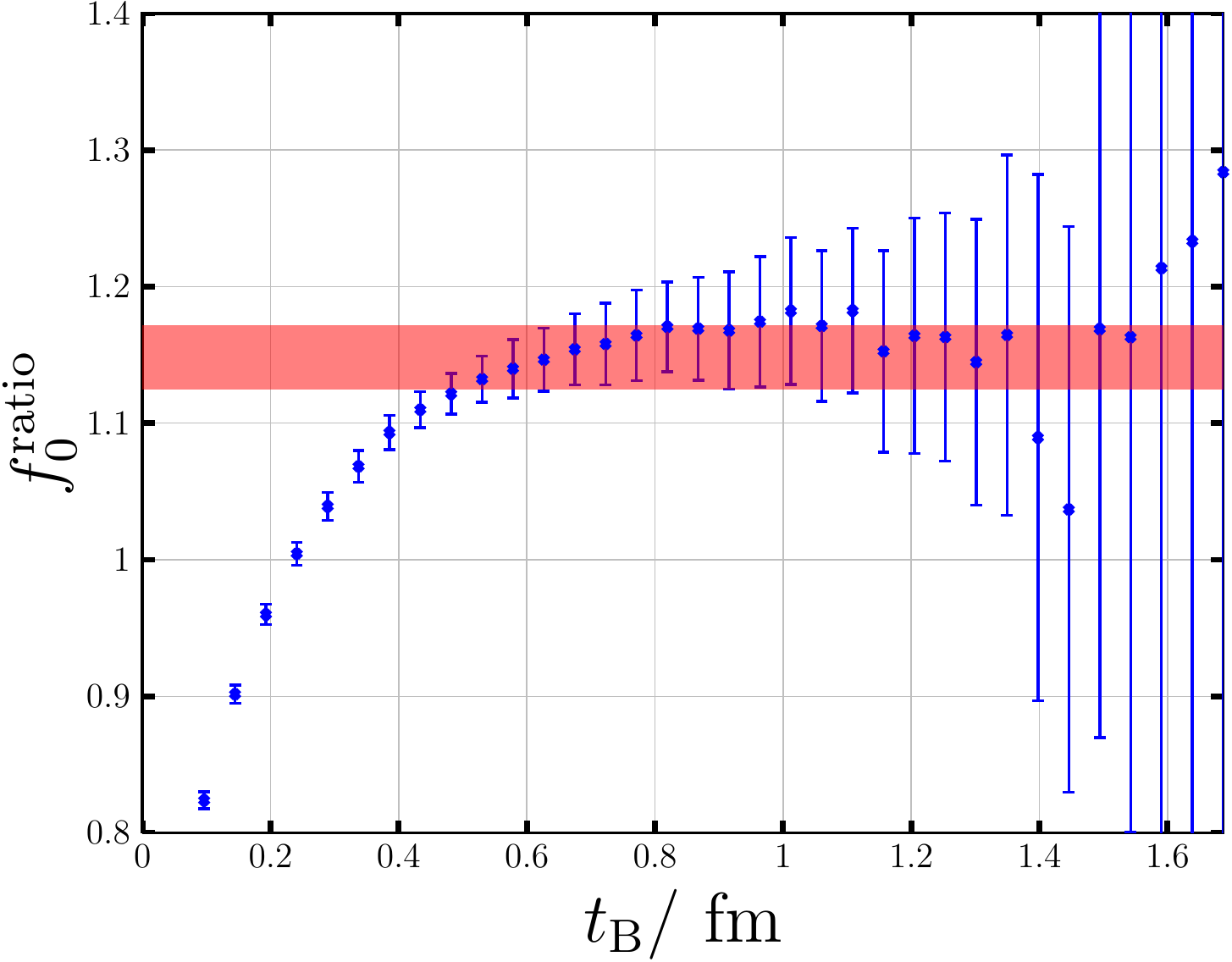}
  \caption{The ratio of the three-point function of a vector current with the pseudoscalar
    interpolating operators of a $B_s$ and a $K$ meson over the square roots of the corresponding
    two-point functions (blue points) as a function of the temporal distance between
    the static vector current and the interpolating operator of the $B_s$ meson.
    The red band is the result of a fit. By courtesy of Ref.~\cite{DellaMorte:2015yda}.}
\label{Fig:semilept}
\end{figure}

\subsection{Static-light correlation functions}
For a static-light two-point correlation function $C_{B}$, the
signal-to-noise ratio squared goes as 
\be
\frac{C^2_{B}(y_0,x_0)}{\sigma^2_{B}(y_0,x_0)} \propto e^{-2 (E_{\rm stat} - M_\pi/2 )|y_0-x_0|}\; ,
\ee
where $E_{\rm stat}$ is the ground-state energy of the $B$-meson which
diverges linearly with the inverse of the lattice spacing. This degradation
is the bottleneck in the computation of the leptonic decay
constant of the $B$-meson in the static limit~\cite{Bernardoni:2014fva}.

The three-point functions needed for the semileptonic decays
$B\!\!\rightarrow\!\!\pi(K) l\nu$, $B\!\!\rightarrow\!\! K (K^*) l l$, etc. have as basic building
blocks the propagators of a static-light meson on one side and of a relativistic meson
with a (large) momentum on the other side.
As we have seen, both of them suffer from an exponential degradation of the
signal-to-noise ratio. At present this sets the limit for the computation of these three-point functions,
and prevent us from determining the form factors at small invariant lepton masses $Q^2$. An example of
the difficulties encountered is shown in Fig.~\ref{Fig:semilept}. There the ratio of the three-point
function of a vector current with the pseudoscalar interpolating operators of a $B_s$ and a $K$ meson
over the square roots of the corresponding two-point functions at  approximatively $Q^2=20$~GeV$^2$
is plotted~\cite{DellaMorte:2015yda}.

\begin{figure}[!thb]
\centering
  \includegraphics[width=0.9\columnwidth]{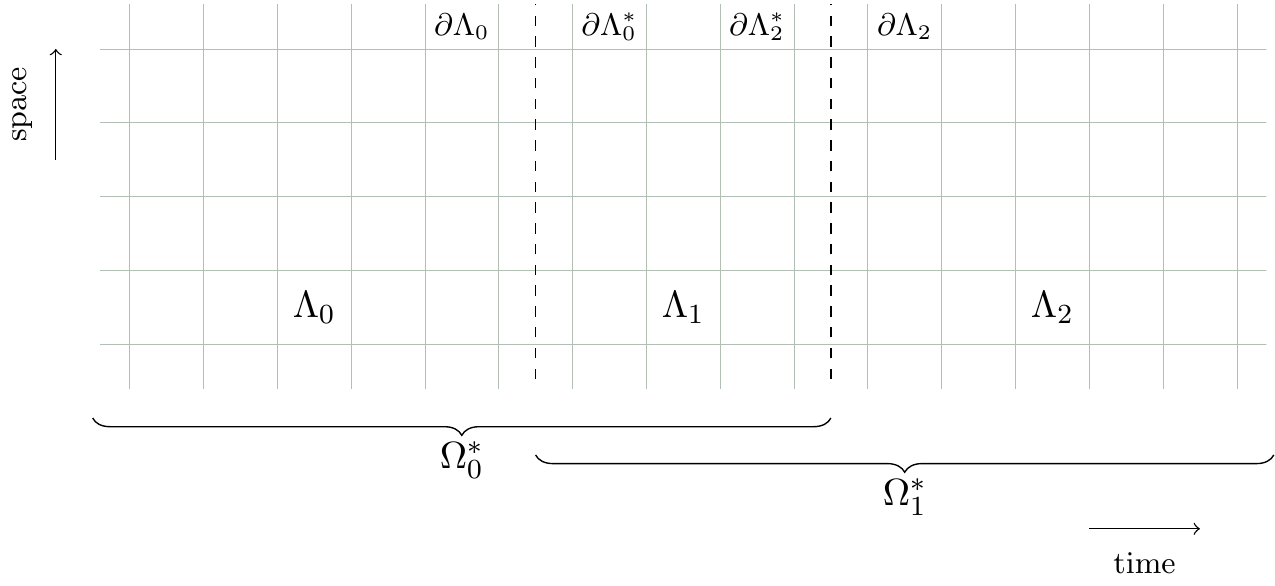}
  \caption{Examples of non-overlapping and overlapping domain decompositions considered in these proceedings.}
  \label{Fig:blocking1}
\end{figure}

\section{Domain decomposition preliminaries}
To find a solution to the signal-to-noise problem, 
we will start by making use of several decompositions of the global lattice in
non-overlapping and overlapping domains~\cite{Luscher:2003qa,Luscher:2005rx}. Without
loss of generality, we will consider a
lattice with periodic and open boundary conditions in the space and time directions,
respectively~\cite{Luscher:2011kk}. The first decomposition of the lattice is in three non-overlapping thick time-slices
$\Lambda_i$, $i=0,1,2$, with the inner and outer (time-slice) boundaries indicated by
$\partial \Lambda_i$ and $\partial \Lambda^*_i$ respectively, see Fig.~\ref{Fig:blocking1}.
It is useful to define projection operators onto the subspaces
of quark fields supported on the domains $\Lambda_i$ as
\be\label{eq:Prj}
   [P_{\Lambda_{i}}\psi](x) =
\begin{cases}
\psi(x) &  x\in \Lambda_i\, , \\
0 & {\rm elsewhere}\; ,
\end{cases}
\ee
and analogously for $P_{\partial \Lambda_{i}}$ and $P_{\partial \Lambda^*_{i}}$.
In the rest of these proceedings we will use the above symbols irrespectively of the
dimension of the full space on which the projectors act.
The Hermitian $O(a)$-improved massive Wilson-Dirac operator
can then be written in the block form\footnote{To keep the notation compact, a block matrix $Q_{\Lambda_{i,j}}$
denotes either a single block of the matrix, or the full matrix with just
that block different from zero. For more details on the notation
used see Ref.~\cite{Ce:2016ajy}}
\begin{equation}
Q=\left ( 
\begin{matrix} 
Q_{\Lambda_{0,0}} & Q_{\Lambda_{0,1}}  & 0      \\
Q_{\Lambda_{1,0}} & Q_{\Lambda_{1,1}}  & Q_{\Lambda_{1,2}} \\
 0     & Q_{\Lambda_{2,1}}  & Q_{\Lambda_{2,2}} \\
\end{matrix}
\right )\; . 
\end{equation}
Maybe the simplest decomposition of the lattice in overlapping
domains is obtained by defining $\Omega^*_i=\Lambda_i\cup\Lambda_{i+1}$ with $i=0,1$,
see Fig.~\ref{Fig:blocking1}. Projection operators on those domains and their boundaries
can be defined analogously to Eq.~(\ref{eq:Prj}). In each of these domains, 
the operator $Q$ takes the block form 
\begin{equation}
Q_{\Omega^*_i}=\left ( 
\begin{matrix} 
Q_{\Lambda_{i,i}} & Q_{\Lambda_{i,i+1}}  \\
Q_{\Lambda_{i+1,i}} & Q_{\Lambda_{i+1,i+1}} 
\end{matrix}
\right )\; .
\end{equation}
\begin{figure}[!thb]
  \centering
  \includegraphics[width=0.9\columnwidth]{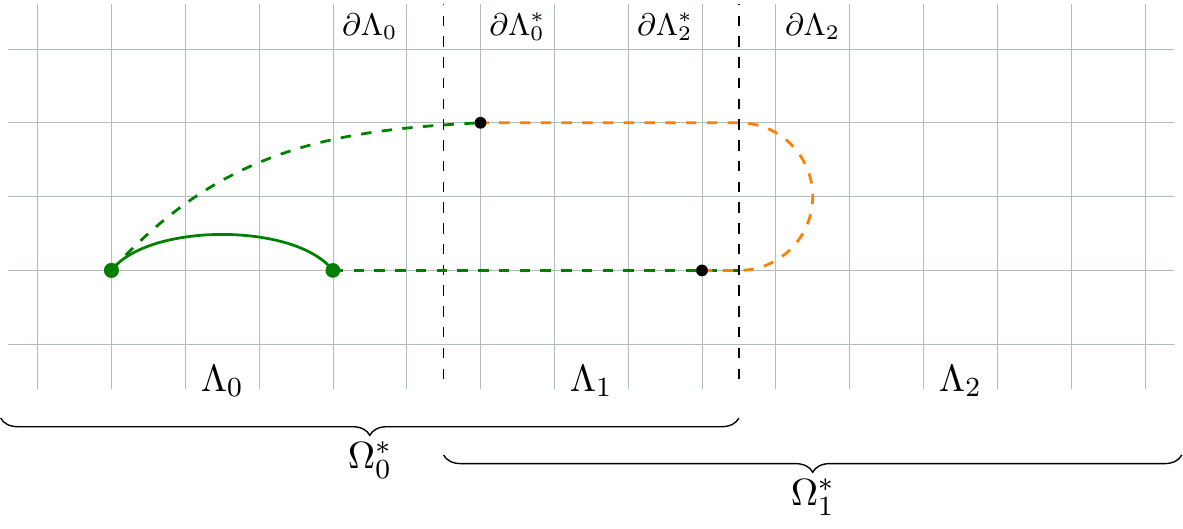}
  \caption{Continuous and dashed lines represent the contributions to the quark propagator as defined
    in Eq.~(\ref{eq:bella}).}
  \label{Fig:blocking1bcp}
\end{figure}
Another domain decomposition which turns out to be instrumental in the following is
the two-block non-overlapping partitioning of the lattice with $\Gamma=\Lambda_0\cup\Lambda_2$
and $\Gamma^*=\Lambda_1$. Notice that $\Gamma$ is a disconnected domain.

\section{Quark propagator and locality}
The quark propagator between two points $x$ and $y$, or better $Q^{-1}(y,x)$,
is formally a non-local functional of the background gauge field over the entire
lattice. Our intuition, however, suggests that $Q^{-1}(y,x)$ should depend weakly
on the values that the gauge field takes far away from the region between $x$ and
$y$. To formalize this insight, we consider the case of $x,y\in\Lambda_0$,
decompose the lattice in the two non-overlapping blocks $\Gamma=\Lambda_0\cup\Lambda_2$
and $\Gamma^*=\Lambda_1$, and choose the thickness $\Delta$ of $\Lambda_1$ so
that $M_\pi \Delta\gg 1$. The Schur complement, defined as usual as
\be
S_{\Gamma} = Q_{\Gamma} - Q_{\partial\Gamma}\, Q_{\Gamma^*}^{-1} \, Q_{\partial\Gamma^*}\; ,
\ee
is then given by 
\begin{equation}
\begin{split}\label{eq:step1}
S_{\Gamma} &=\left ( 
\begin{matrix} 
Q_{\Lambda_{0,0}}-Q_{\Lambda_{0,1}} Q_{\Lambda_{1,1}}^{-1} Q_{\Lambda_{1,0}} & -Q_{\Lambda_{0,1}} Q_{\Lambda_{1,1}}^{-1}
Q_{\Lambda_{1,2}} \\[0.25cm]
-Q_{\Lambda_{2,1}} Q_{\Lambda_{1,1}}^{-1} Q_{\Lambda_{1,0}}  & Q_{\Lambda_{2,2}} -Q_{\Lambda_{2,1}} Q_{\Lambda_{1,1}}^{-1} Q_{\Lambda_{1,2}} 
\end{matrix}
\right ) \,.
\end{split}
\end{equation}
By noticing that
\be\label{eq:ssc}
P_{\Lambda_{0}}\, Q^{-1}_{\Omega^*_0}\, P_{\Lambda_{0}} =
\left[ Q_{\Lambda_{0,0}}-Q_{\Lambda_{0,1}} Q_{\Lambda_{1,1}}^{-1} Q_{\Lambda_{1,0}}\right]^{-1}\; , \qquad
P_{\Lambda_{2}}\, Q^{-1}_{\Omega^*_1}\, P_{\Lambda_{2}} =
\left[ Q_{\Lambda_{2,2}}-Q_{\Lambda_{2,1}} Q_{\Lambda_{1,1}}^{-1} Q_{\Lambda_{1,2}}\right]^{-1}\; , 
\ee
after a few steps of algebra one obtains
\be\label{eq:bella}
P_{\Lambda_{0}}\, Q^{-1}\, P_{\Lambda_{0}} = P_{\Lambda_{0}} \left\{Q_{\Omega_0^*}^{-1} +
Q^{-1}_{\Omega_0^*} Q_{\Lambda_{1,2}} Q^{-1}_{\Omega_1^*}  Q_{\Lambda_{1,0}}\, \frac{1}{1-w}\, Q^{-1}_{\Omega_0^*}
\right\} P_{\Lambda_{0}}\; , 
\ee
where
\be\label{eq:w}
w=P_{\partial\Lambda_{0}} Q^{-1}_{\Omega^*_0}\, Q_{\Lambda_{1,2}}\, Q^{-1}_{\Omega^*_1} Q_{\Lambda_{1,0}} \, .
\ee
\begin{figure}[thb]
  \centering
  \includegraphics[width=0.9\columnwidth]{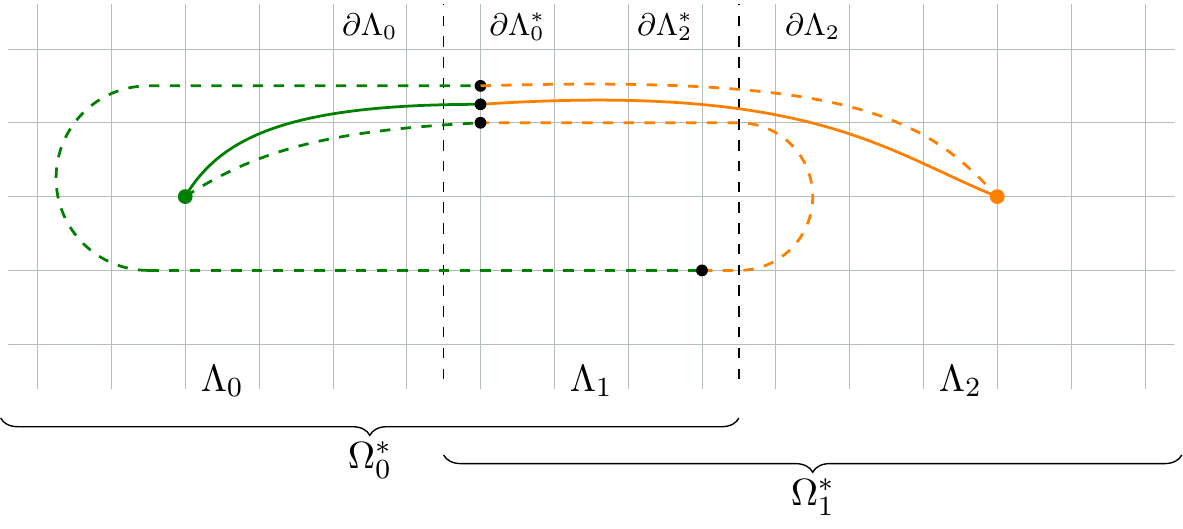}
  \caption{Continuous and dashed lines represent the first two contributions to the quark propagator
    in Eq.~(\ref{eq:SAP}), corresponding to $n=0$ and $1$.}
  \label{Fig:blocking3ap}
\end{figure}
Our intuition is formalized in Eq.~(\ref{eq:bella}). The first term on the r.h.s. (continuous line in
Fig.~\ref{Fig:blocking1bcp}) does not depend on the gauge field in $\Lambda_2$.
That dependence comes only from the second term (dashed line in Fig.~\ref{Fig:blocking1bcp})
which propagates a quark from $x\in \Lambda_0$ to the region $\Lambda_2$ and
back to $y\in \Lambda_0$. The contribution from these paths is suppressed
proportionally to $e^{-M_\pi\Delta}$ thanks to Eq.~(\ref{eq:confbyconf}).

The very same domain decomposition sheds light also on the gauge-field dependence of $Q^{-1}(y,x)$
when $x$ and $y$ are in distant blocks, e.g. $x\in\Lambda_0$ and $y\in\Lambda_2$. By
following an analogous derivation, one arrives to
\be\label{eq:bella2}
P_{\Lambda_{2}}\, Q^{-1}\, P_{\Lambda_{0}} = -
P_{\Lambda_{2}}\, Q_{\Omega_1^*}^{-1} Q_{\Lambda_{1,0}}\, \frac{1}{1-w}\, Q_{\Omega_0^*}^{-1} P_{\Lambda_{0}} \; .
\ee
From Eq.~(\ref{eq:w}) it is clear that the operator $w$ propagates a
quark from the inner boundary of $\Lambda_0$ to $\Lambda_2$ and back to $\partial \Lambda_0$, and
it is therefore suppressed proportionally to $e^{-M_\pi\Delta}$. By neglecting its contribution on the
r.h.s of Eq.~(\ref{eq:bella2}), the bulk of the quark propagator turns out to have a factorized 
dependence on the gauge field in $\Lambda_0$ and $\Lambda_2$. 

The Eqs.~(\ref{eq:bella}) and (\ref{eq:bella2}) provide further insight into the
dependence of the quark propagator from the gauge field. In Eq.~(\ref{eq:bella2}),
for instance, we can expand the factor $(1-w)^{-1}$ into the Neumann series to obtain
\be\label{eq:SAP}
P_{\Lambda_{2}}\, Q^{-1}\, P_{\Lambda_{0}} = -
P_{\Lambda_{2}}\, Q_{\Omega_1^*}^{-1} Q_{\Lambda_{1,0}}
\sum_{n=0}^{\infty} \left[Q^{-1}_{\Omega^*_0}\, Q_{\Lambda_{1,2}}\, Q^{-1}_{\Omega^*_1} Q_{\Lambda_{1,0}}\right]^n
Q_{\Omega_0^*}^{-1}\, P_{\Lambda_{0}} \; .
\ee
We recognize on the r.h.s the result of a Schwarz alternating procedure (SAP) with overlapping
domains $\Omega^*_0$ and $\Omega^*_1$. The propagator is written as a
series of terms, each having a factorized gauge-field dependence of increasing complexity.
The index $n$ counts the number of times a quark loops from the inner boundary of 
$\Lambda_0$ to $\partial \Lambda_2$ and back to $\partial \Lambda_0$  before arriving in $y$.
The contribution of these paths is suppressed proportionally to
$e^{-n M_\pi\Delta}$. The thickness $\Delta$ of the overlapping region regulates the
rate of convergence of the associated Neumann series, making SAP with overlapping
domains also a valid alternative for computing the quark propagator in lattice QCD
with respect to the case of non-overlapping domains\footnote{The suppression of the
quark propagator already after $0.5$~fm or so may be enough for this procedure and the
factorization of the determinant in the next section to work in practice.}~\cite{Luscher:2003qa}.

\section{Block decomposition of the determinant\label{sec:BD}}
The domain decomposition of the lattice in the two blocks $\Gamma=\Lambda_0\cup\Lambda_2$
and $\Gamma^*=\Lambda_1$ is also the starting point for the factorization of the gauge-field
dependence of the quark determinant. The LU decomposition of the associated $2$ by $2$ block
form of the Dirac operator leads to 
\begin{equation}
\begin{split}
\hspace{-0.5cm}\det\, Q&=\det Q_{\Lambda_{1,1}} \det\, \left ( 
\begin{matrix} 
Q_{\Lambda_{0,0}}-Q_{\Lambda_{0,1}} Q_{\Lambda_{1,1}}^{-1} Q_{\Lambda_{1,0}} & -Q_{\Lambda_{0,1}} Q_{\Lambda_{1,1}}^{-1}
Q_{\Lambda_{1,2}} \\[0.25cm]
-Q_{\Lambda_{2,1}} Q_{\Lambda_{1,1}}^{-1} Q_{\Lambda_{1,0}}  & Q_{\Lambda_{2,2}} -Q_{\Lambda_{2,1}} Q_{\Lambda_{1,1}}^{-1} Q_{\Lambda_{1,2}} 
\end{matrix}
\right ) \,,
\end{split}
\end{equation}
which, thanks to Eqs.~(\ref{eq:ssc}), can be re-written as 
\begin{equation}
  \det\, Q= \frac{1}{\det\, Q^{-1}_{\Lambda_{1,1}}
    \det\left[P_{\Lambda_{0}}\,  Q^{-1}_{\Omega^*_0} \, P_{\Lambda_{0}}\right]
    \det\left[P_{\Lambda_{2}}\,  Q^{-1}_{\Omega^*_1} \, P_{\Lambda_{2}}\right]}
  \det\,
\left (
\begin{matrix} 
1 & P_{\Lambda_{0}} Q^{-1}_{\Omega^*_0} Q_{\Lambda_{1,2}} \\[0.25cm]
P_{\Lambda_{2}} Q^{-1}_{\Omega^*_1} Q_{\Lambda_{1,0}}  & 1
\end{matrix}
\right )\; . 
\end{equation}
By noticing that the last determinant on the r.h.s can be reduced to the one of a
matrix acting on one of the boundaries only, it is easy to show that
\begin{equation}\label{eq:detw}
\hspace{-0.8cm}\det\,
\left (
\begin{matrix} 
1 & \!\!\!\!\!\!P_{\Lambda_{0}}\, Q^{-1}_{\Omega^*_0} Q_{\Lambda_{1,2}} \\[0.25cm]
P_{\Lambda_{2}}\, Q^{-1}_{\Omega^*_1} Q_{\Lambda_{1,0}}  & \!\!\!\!\!\!1
\end{matrix}
\right )
= \det\, (1 - w)\, ,\\[0.375cm]
\end{equation}
and therefore 
\begin{equation}\label{eq:factfinal}
  \det\, Q= \frac{1}{\det\, Q^{-1}_{\Lambda_{1,1}}
    \det\left[P_{\Lambda_{0}}\,  Q^{-1}_{\Omega^*_0} \, P_{\Lambda_{0}}\right]
    \det\left[P_{\Lambda_{2}}\,  Q^{-1}_{\Omega^*_1} \, P_{\Lambda_{2}}\right]
    \det\, \left[1-w\right]^{-1}} \; .  
\end{equation}
For the first three determinants on the r.h.s, the goal has been reached:
$\det Q^{-1}_{11}$ depends on the gauge field in the block $\Lambda_1$,
$\det\, [P_{\Lambda_{0}}\,  Q^{-1}_{\Omega^*_0} \, P_{\Lambda_{0}}]$ on the gauge field in
$\Omega^*_0$, and $\det\, [P_{\Lambda_{2}}\,  Q^{-1}_{\Omega^*_1} \, P_{\Lambda_{2}}]$
on the gauge field in $\Omega^*_1$. The (small) remaining determinant $\det\, [1-w]$
still depends on the gauge field over the whole lattice. As in Eq.~(\ref{eq:SAP}),
we can expand the factor $[1-w]^{-1}$ into the Neumann series and obtain
\be\label{eq:MB}
\frac{1}{\det\, [1-w]^{-1}} = \frac{\det\,[1-R_{N+1}(1-w)]}{\det\,\left[\sum_{k=0}^{N} w^k\right]} =
\frac{\det\,[1-R_{N+1}(1-w)]}{\prod_{k=1}^{N/2} {\det} \big[ (u_k -w )^\dagger (u_k -w) \big]}\, ,
\ee
where $N$ is chosen to be even, $u_k=e^{i\frac{2\pi k}{N+1}}$ ($k=1,\dots,N$) are the roots of the approximant
polynomial $\sum_{k=0}^N w^k$, and the remainder polynomial is $R_{N+1}(1-w)=w^{N+1}$.
For the last equality we have used the fact that the roots of the approximant polynomial 
come in complex conjugate pairs and that $w$ is similar to $w^\dagger$ \cite{Ce:2016ajy}.
We recognize in Eq.~(\ref{eq:MB}) a specific implementation of L\"uscher's original multi-boson
proposal~\cite{Luscher:1993xx} generalized to complex matrices~\cite{Borici:1995np,Borici:1995bk,Jegerlehner:1995wb},
see Ref.~\cite{Ce:2016ajy} for a more general discussion. By defining the matrix 
\be\label{eq:Wz}
W_z = \begin{pmatrix} 
 z \, P_{\partial\Lambda_{0}} & P_{\partial\Lambda_{0}} Q^{-1}_{\Omega^*_0} Q_{\Lambda_{1,2}}  \\[0.25cm]
  P_{\partial\Lambda_{2}} Q^{-1}_{\Omega^*_1} Q_{\Lambda_{1,0}}  & z \, P_{\partial\Lambda_{2}} 
  \end{pmatrix}\; , 
\ee
we can perform the reverse substitution of the one in Eq.~(\ref{eq:detw}), and finally obtain
\begin{equation}
  \prod_{k=1}^{N/2} {\det} \big[ (u_k -w )^\dagger (u_k -w) \big]  =
  \prod_{k=1}^{N/2} {\det} \big[ W_{\sqrt{u_k}}^\dagger \, W_{\sqrt{u_k}} \big]\; . 
\label{e:pV}
\end{equation}
It is this expression with $W_z$ acting on $\partial\Lambda_{0}$ and  $\partial\Lambda_{2}$ that
will allow in the next section for a fully factorized domain decomposition of the fermion action.
For the determination of the approximation, however, it has been advantageous to work with the
operator $w$ (acting on $\partial\Lambda_{0}$ only) since the order of the polynomial $N$ is reduced
by about a factor of $2$ for a given accuracy. Notice that the multi-boson contribution in
Eq.~(\ref{e:pV}) is manifestly positive for each single flavour.

\section{Multi-level integration with fermions}
By introducing auxiliary pseudofermion and multi-boson fields, for two flavors of quarks we can
finally represent the
determinants in Eqs.~(\ref{eq:factfinal}) and (\ref{e:pV}) as\footnote{The identity
$\det Q^{-1}_{\Lambda_{1,1}} \cdot \det\,  [P_{\Lambda_{0}} Q^{-1}_{\Omega^*_0} P_{\Lambda_{0}}] =
\det\, Q^{-1}_{\Omega^*_0}$ can be used to speed up the simulation when region $1$ is active.}
\begin{equation}\label{eq:act}
\begin{split}
& \frac{\det Q^2}{\det\{1-R_{N+1}(1-w)\}^2} =\frac{1}{
    \det\, [Q_{\Lambda_{1,1}}^{-1}]^2 \cdot \det\,  [P_{\Lambda_{0}} Q^{-1}_{\Omega^*_0} P_{\Lambda_{0}}]^{2} \cdot
\det\, [P_{\Lambda_{2}}  Q^{-1}_{\Omega^*_1}P_{\Lambda_{2}}]^{2} } \times\\[0.25cm]
  & \times \frac{1}{\prod_{k=1}^{N} {\det} \big[ W_{\sqrt{u_k}}^\dagger \, W_{\sqrt{u_k}} \big]} =
  \, C' \int [d\phi_0d\phi_0^\dagger]\, e^{-|P_{\Lambda_{0}}  Q^{-1}_{\Omega^*_0} \phi_0|^2 }
 \int [d\phi_1d\phi_1^\dagger]\, e^{-|Q_{\Lambda_{1,1}}^{-1}\phi_1|^2 }\cdot\\
      & \int [d\phi_2 d\phi_2^\dagger]\, e^{-|P_{\Lambda_{2}}  Q^{-1}_{\Omega^*_1} \phi_2|^2 }\cdot 
       \prod_{k=1}^{N} \left \{
       \int [d\chi_k d\chi_k^\dagger] e^{-|W_{\sqrt{u_k}} \chi_k|^2 } \right \}\; , \\
\end{split}
\end{equation}
where $C'$ is an irrelevant numerical constant.
Each pseudofermion field $\phi_i$ is confined to the corresponding region $\Lambda_i$, $i=0,1,2$.
The $N$ multi-boson fields $\chi_k$ live on the outer boundaries of region $\Lambda_1$. We can
decompose them as $\chi_k=\eta_k+\xi_k$, with
$\eta_k=P_{\partial\Lambda_{0}} \chi_k$ and $\xi_k=P_{\partial\Lambda_{2}} \chi_k$, and split explicitly the
contributions from the inner boundaries of regions $\Lambda_0$ and $\Lambda_2$ as 
\begin{equation}
\label{eq:multiboson}
\begin{split}
|W_{z} \chi_k|^2 &=
|z|^2 |\eta_k|^2 + |z|^2 |\xi_k|^2+|P_{\partial\Lambda_{2}} Q^{-1}_{\Omega^*_1} Q_{\Lambda_{1,0}} \eta_k|^2+|P_{\partial\Lambda_{0}} Q^{-1}_{\Omega^*_0}
Q_{\Lambda_{1,2}} \xi_k|^2\\
&+\big[z (\xi_k,Q_{\Lambda_{2,1}} Q^{-1}_{\Omega^*_0} \eta_k) +z^*\, (\xi_k, Q^{-1}_{\Omega^*_1} Q_{\Lambda_{1,0}} \eta_k)
  + \text{c.c.}\big]\; .
\end{split}
\end{equation}
{\it The dependence of the bosonic action from the gauge field in block $\Lambda_0$ and
$\Lambda_2$ is thus factorized}. Interestingly, the terms in Eq.~(\ref{eq:multiboson})
which contribute to the forces in region $\Lambda_0$ always start (or end) on the inner
boundary of $\Lambda_2$ and vice versa. The matrices in Eq.~(\ref{eq:multiboson})
contain one boundary to boundary propagator which is suppressed exponentially in $\Delta$,
and so are the corresponding forces.

The factorization of the gauge-field dependence in the bosonic action has been achieved
by treating differently the contributions from the various quark paths to the
fermion determinant. Those with no loops around the inner boundaries of
$\Lambda_0$ and $\Lambda_2$ have a factorized dependence on the gauge field in $\Lambda_0$
and $\Lambda_2$, and can then be included by introducing the pseudofermion fields
$\phi_i$ in each of the three blocks. The contributions from quark paths with 1 up to
$N$ loops around $\partial \Lambda_0$ and $\partial \Lambda_2$ are introduced via
the multi-boson fields living on these boundaries and their interactions. The contributions from
higher loops are either negligible within the precision required, or can be associated to the
observables in the form of a reweighting factor, see below.

We are now in the position to formulate a multi-level numerical integration for lattice QCD.
A given correlation function of a string of fields $O$ can be written as 
\be\label{eq:rwgt}
\langle O \rangle  = \frac{\langle O\, {\cal W}_N \rangle_N}{\langle {\cal W}_N \rangle_N}
= \langle O^{\rm fact}\, \rangle_N +
\frac{\langle O\,  {\cal W}_N - O^{\rm fact}\, \langle {\cal W}_N \rangle_N\; \rangle_N}{\langle {\cal W}_N \rangle_N}\; ,
\ee
where $O^{\rm fact}$ is a rather precise factorized approximation of
$O$ that can be obtained by expressing the quark propagators in the fermionic Wick contractions following
Eqs.~(\ref{eq:bella}) and (\ref{eq:bella2}) as in Ref.~\cite{Ce:2016idq}, and $\langle\cdot\rangle_N$
indicates the
expectation value in the theory defined by $N$ multi-boson fields. Since both the action and the observable are
factorized, the expectation value $\langle O^{\rm fact}\, \rangle_N$ can be computed with
a multi-level algorithm by generating gauge field configurations with the multi-boson action at
finite $N$. All other quantities in Eq.~($\ref{eq:rwgt}$) can be computed with a standard one-level
Monte Carlo procedure. For two flavors, the reweighting factor ${\cal W}_N$ is
\be
{\cal W}_N = \det\{1-R_{N+1}(1-w)\}^2\,. 
\ee
This expression is easily evaluated as 
\be\label{eq:WNeta}
{\cal W}_N = \frac{\int [d \eta] [d\eta^\dagger]
e^{-|(1-R_{N+1})^{-1} \eta|^2}}{\int [d \eta] [d\eta^\dagger] e^{-\eta^\dagger \eta}}\; ,
\ee
where the exponent can be computed by a Taylor expansion,
and as usual the integral over $\eta$ can be replaced by random samples.

\begin{figure}[t!]
  \centering
  \includegraphics[width=0.49\columnwidth]{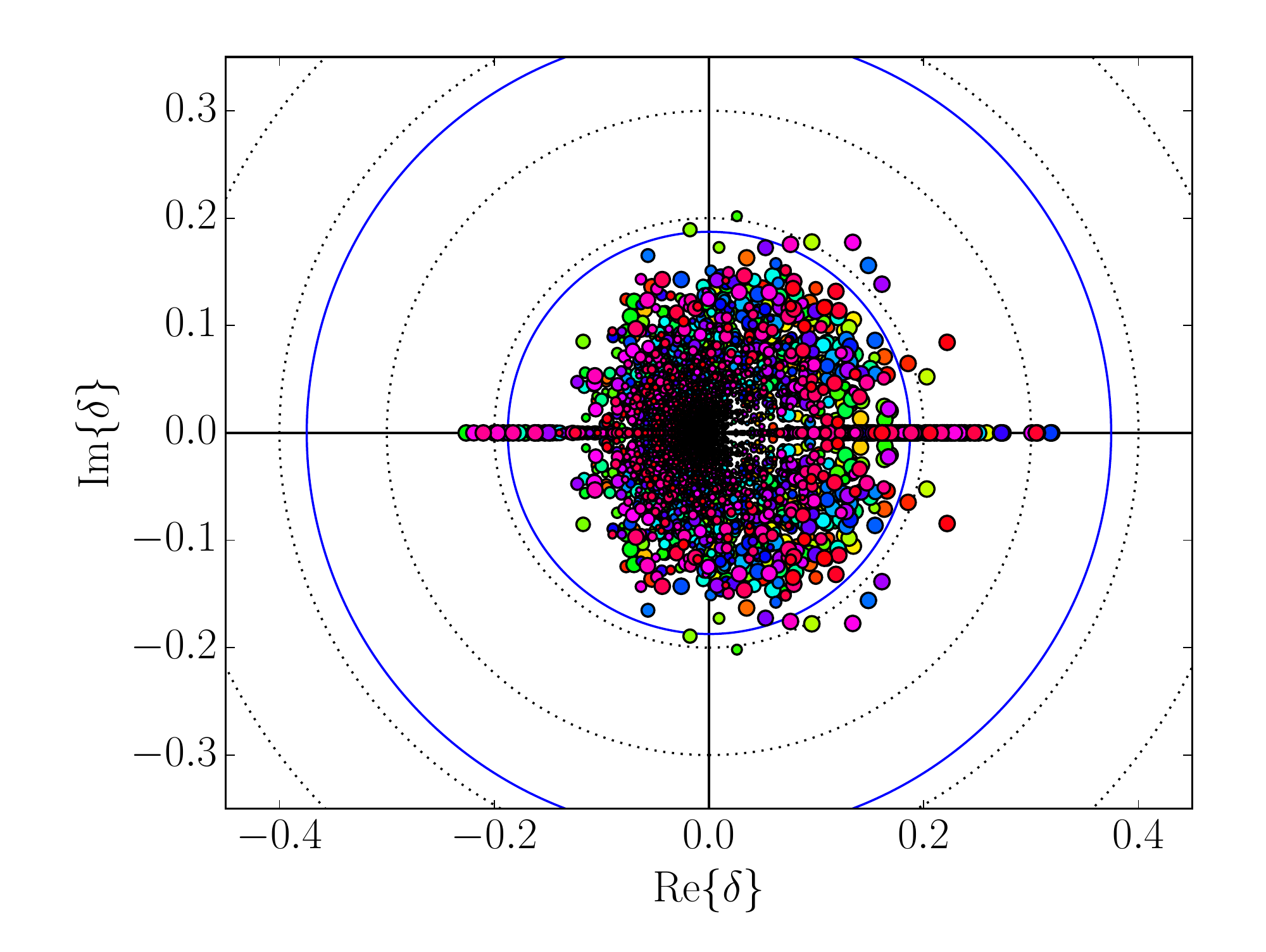}
  \includegraphics[width=0.49\columnwidth]{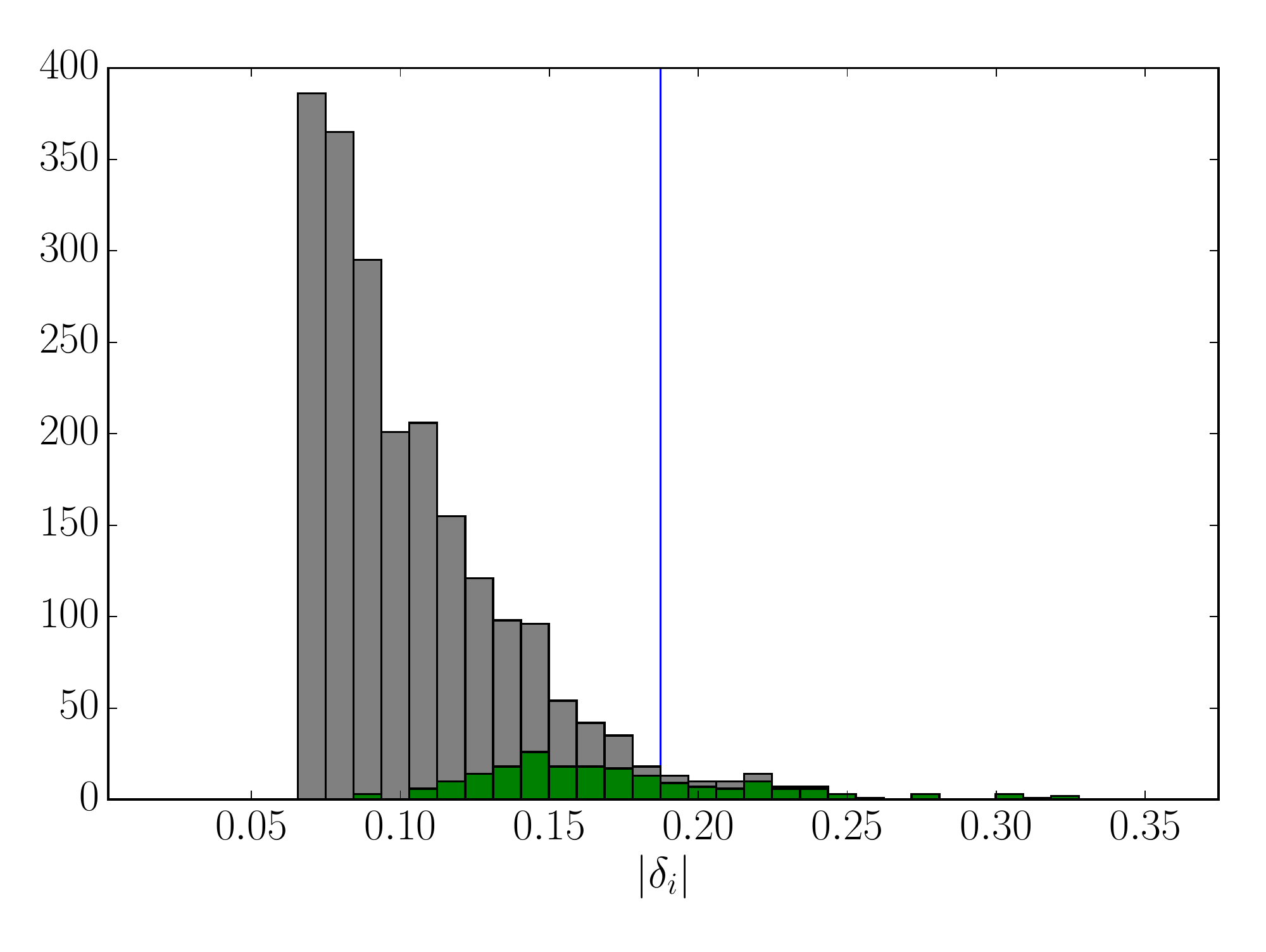}
  
  \caption{Left: the 60 largest eigenvalues $\delta_i$ of $w$ ($\Delta=12\, a$) for
    $200$ configurations; the blue circles have radius
    $\bar\delta=\exp{-M_\pi \Delta}$ and $2\, \bar\delta$.
    Right: distributions of the eigenvalues $\delta_i$ of $w$ with the
    largest absolute norm (green) and $|\delta_i|>0.35\,\bar{\delta}$ (grey);
    the vertical blue line is at $\abs{\delta_i}=\bar\delta$.}
\label{fig:spectrum_2cfg}
\end{figure}

\section{A crucial numerical test\label{s:numw}}
The feasibility of the whole proposal hinges crucially on the assumption that the spectrum of
the operator $(1-w)$ is confined into a disk around $1$ in the complex plane, with a radius
significantly below unity. Only in this case, a small number of bosonic fields $N$ in
Eq.~(\ref{eq:multiboson}) leads to a good enough approximation at a reasonable computational cost.

To test this assumption, $200$ configurations 
with the Wilson gluonic action and with two flavors of nonperturbatively $O(a)$-improved Wilson quarks have been
generated in Ref.~\cite{Ce:2016ajy}, with $\beta=6/g_0^2=5.3$, $T\times L^3=64\times 32^3 a^4$ and open boundary conditions.
The lattice spacing is $a=0.0652(6)\,\fm$, while the pion mass is
$a M_\pi=0.1454(5)$ corresponding to $440(5)$~MeV, see Ref.~\cite{Ce:2016ajy} for more details.

For $\Delta/a=8,12$ and $16$, $60$ approximate eigenvalues
$\delta_i$ of $w$ with the largest absolute value have been computed with the Arnoldi algorithm.
In the left plot of Fig.~\ref{fig:spectrum_2cfg}
all eigenvalues for all $200$ configurations are shown for $\Delta=12\, a$. As expected,
they are either real or appear in complex conjugate pairs. The blue circles in these plots
have radius $\bar \delta$ and $2\, \bar\delta$, where $\bar \delta = \exp{-M_\pi \Delta}$.
\begin{table}[t]
  \centering
  \caption{Properties of the spectrum of $w$ for different values of $\Delta$.}
\label{tab:spectrum}
\begin{tabular}{S[table-format=2]S[table-format=1.4]S[table-format=1.4]S[table-format=1.4]S[table-format=1.4]}
    \toprule
{$\Delta/a$} & {$\bar\delta$} & {$\ev{\max_i\abs{\delta_i}}$} &
{$\sigma(\max_i\abs{\delta_i})$} &
 {$\max\max_i\abs{\delta_i}$} \\
    \midrule
        8 & 0.3273 & 0.2886 & 0.0616 & 0.5130 \\
       12 & 0.1710 & 0.1692 & 0.0453 & 0.3193 \\
       16 & 0.1072 & 0.0951 & 0.0284 & 0.1977 \\
    \bottomrule
  \end{tabular}
\end{table}
The distribution of the eigenvalue with the largest magnitude is
shown in green in the right plot of Fig.~\ref{fig:spectrum_2cfg}. It is peaked at a value slightly
smaller than $\bar\delta$, denoted by a vertical blue line, and extends up to $\approx 2\bar\delta$.
The results for the largest eigenvalue norm computed over the $200$ configurations, its
average value and the estimate of its standard deviation
are also reported in Table~\ref{tab:spectrum}. In the right plot of Fig.~\ref{fig:spectrum_2cfg}
we also report in grey the distribution of the absolute value of the eigenvalues limited to those
with $\abs{\delta_i}>0.35\,\bar{\delta}$. 

A clear picture emerges from these data. The largest eigenvalue of the relevant operator $w$ decreases
proportionally to $\exp{-M_\pi \Delta}$ in this range of values of $\Delta$,
with a prefactor of order $1$. This in turn implies that
$(1-w)$ has a large gap if $\Delta$ is properly tuned. The relative error on the determinant
at various values of $N$ compares well with $|\delta|^{N+1}_{\rm max}$ configuration
by configuration~\cite{Ce:2016ajy}.  No big prefactors appear because the
eigenvalues do not accumulate near the maximum one, and the approximation gets exponentially more
precise toward the center of the circle. The reweighting factor, as defined in Eq.~(\ref{eq:WNeta}) for $N=12$ 
and estimated with 4 random sources per configuration, deviates from $1$ by at most $4.5\cdot 10^{-6}$, again
in line with the expectation. At the level of precision of most contemporary simulations the impact of the
reweighting factor is therefore negligible.

\section{Numerical tests of MB-DD-HMC}
The effective action in \Eq{eq:act} can be simulated by using variants
of the hybrid Monte Carlo algorithm~\cite{Duane:1987de}. The introduction of
multi-boson fields and the resulting multi-boson domain-decomposed
hybrid Monte Carlo (MB-DD-HMC) do not pose particular problems, see
Ref.~\cite{Ce:2016ajy} for more details on its implementation. For a first test of its potentiality,
a subset of $n_0=32$ configurations spaced by at least $80$ molecular dynamics units (MDUs) among
the $200$ described in \Sect{s:numw} has been selected in Ref.~\cite{Ce:2016ajy}. Starting
from each of them, $n_1=45$ level-1 configurations spaced by $4$ MDUs have been generated by keeping
fixed the spatial links on the boundaries $\partial\Lambda_{0}$ and $\partial\Lambda_{2}$  and all the
links in between. The region $\Lambda_1$ extends between time slices $24$ and $35$, corresponding to a
thickness of $\Delta\approx0.8$\,fm and $M_\pi\Delta \approx 1.7$. 
\begin{figure}
\begin{center}
\includegraphics[width=0.48\textwidth,clip]{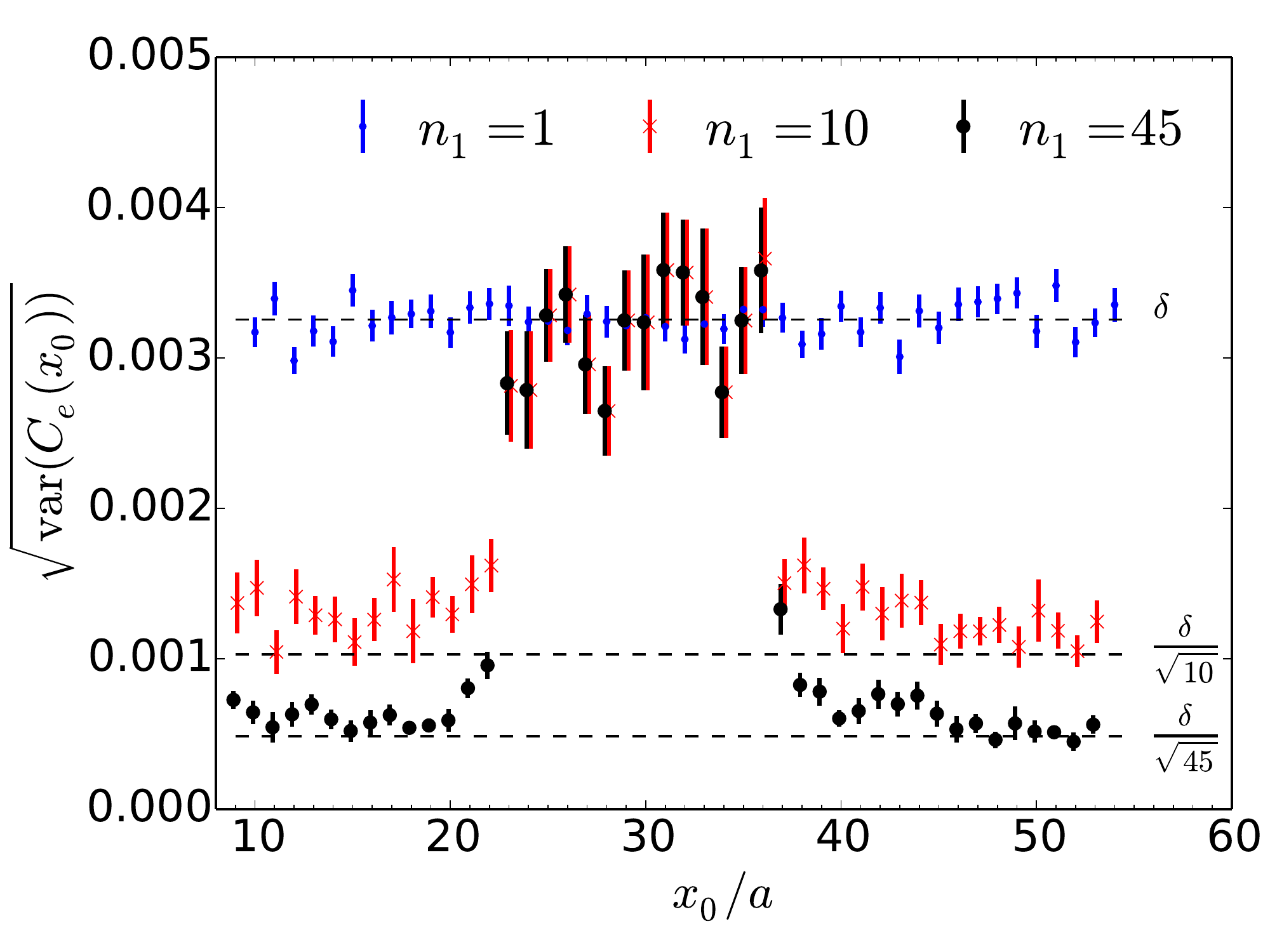}
\hspace{0.02\textwidth}
\includegraphics[width=0.48\textwidth,clip]{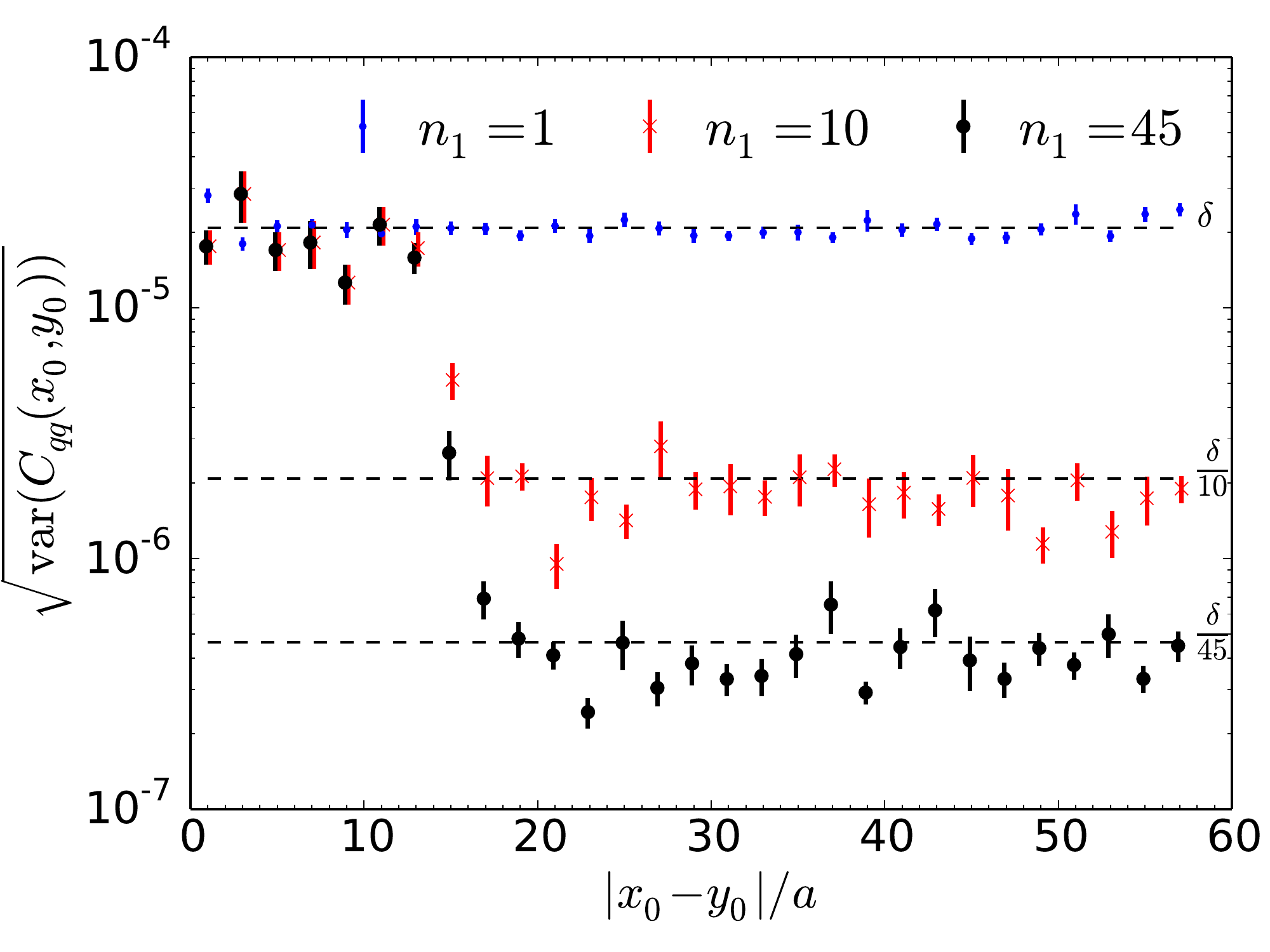}
\end{center}
\caption{\label{f:num}In the left panel, the square root of the variance of the energy density averaged 
over the time slice $x_0$ is shown. In the frozen central region this does not profit from
the level-1 updates, while in the active regions, it decreases with the square root of their inverse number.
The right plot demonstrates the effectiveness of the multi-level algorithm for the topological charge
density correlation function. The time slices $x_0$ and $y_0=30\, a - x_0$ are chosen such that they are
symmetric
with respect to the frozen region $\Lambda_1$. Once $|y_0-x_0|>12\, a$, the densities enter the
active regions where the
square root variance decreases with $1/n_1$. In both plots, the horizontal lines indicate the ideal scaling
behavior as expected from the variance measured at level 0.
}
\end{figure}
Maybe the simplest observables to be computed for a first test of the algorithm are the one-
and two-point gluonic correlation functions
\be
\begin{split}
C_{e} (x_0) = &\frac{1}{L^3} \langle \bar e(x_0) \rangle\; ,\\
C_{ee} (x_0,y_0) = & \frac{1}{L^3} \langle \bar e(x_0)\, \bar e(y_0)  \rangle_c\; , \\
C_{qq} (x_0,y_0) = & \frac{1}{L^3} \langle \bar q(x_0)\, \bar q(y_0)  \rangle\; ,
\end{split}
\ee
of the energy and the topological charge densities summed over the time slices 
\be
\bar e (x_0) = \frac{1}{4} \sum_{\vec x} F_{\mu\nu}^{a}(x)
F_{\mu\nu}^{a}(x)\;, \quad 
\bar q(x_0)  =  \frac{1}{64\pi^2}\, \sum_{\vec x} 
\epsilon_{\mu\nu\rho\sigma}\, F_{\mu\nu}^a(x) F_{\rho\sigma}^a(x)\; ,
\ee
with $F_{\mu\nu}(x)$ being the gluon field strength tensor,
see Ref.~\cite{Ce:2016ajy} for more details.

The two-level estimates of these quantities have been carried out
by first averaging, for each of the $n_0$ configurations, the
densities over the $n_1$ level-1 background fields. This gives 
the $n_0$ measurements of the improved estimator for the one-point function,
while for the two-point correlators the $n_0$ measurements are obtained
by multiplying the improved densities.
The figure of merit is the variance of the estimators. In the situation
where autocorrelations among
the $n_0$ level-0 configurations can be neglected, the square root of the
variance divided by $\sqrt{n_0}$ gives the error of the measurement.
Since the cost of the simulation scales linearly in $n_1$, the variance itself
should decrease with $n_1$ to break even.

The square root of the variance of $C_{e} (x_0)$ as a function of $x_0$ is shown
in the left panel of Fig.~\ref{f:num} for various values of $n_1$.
In the central region, the links are frozen during the level-1 updates.
We therefore do expect the same variance as in the level-0 estimator, but with a
larger error since in this case the number of level-0 configurations is $32$
instead of the $200$ used in the standard case. Once
the density moves into the active regions $\Lambda_0$ and $\Lambda_2$, however, the variance of the 
estimator is clearly improved, in agreement with what is expected from 
ideal scaling, i.e.\ $\sqrt{\mathrm{var}(C_e)}\propto 1/\sqrt{n_1}$.

In the right panel the same analysis is shown for the two-point function
$C_{qq}$, and analogous results are obtained for $C_{ee}$.
Here the full benefit of the method can be realized, because an improved
estimator can be constructed by averaging for each of the $n_0$ fields the
densities in regions $\Lambda_0$ and $\Lambda_2$ 
independently before constructing the two-point function. As optimal
scaling in this case we expect a reduction of the square root of the variance, 
and therefore the error, with $1/n_1$. The numerical data are in agreement
with such a reduction once $x_0$ and $y_0$ are in two different active regions.

These results are in line with expectations. 
In the region where the links are
frozen during the level-1 updates no benefit from the multi-level is observed. As soon
as the densities are in the active regions $\Lambda_0$ and $\Lambda_2$, the square root of
the variances of the one- and two-point functions are reduced by $1/\sqrt{n_1}$ and
$1/n_1$ respectively. The two-level Monte Carlo works at full potentiality in these regions,
with a net gain of a factor $n_1$ in the signal-to-noise ratio of the two-point function.
This in turn
implies that links in the active regions $\Lambda_0$ and $\Lambda_2$ are regularly updated
during the level-1 MB-DD-HMC, and no particular freezing induced by multi-boson fields is
observed.
\begin{figure}[!t]
  \centering
  \includegraphics[width=0.49\columnwidth]{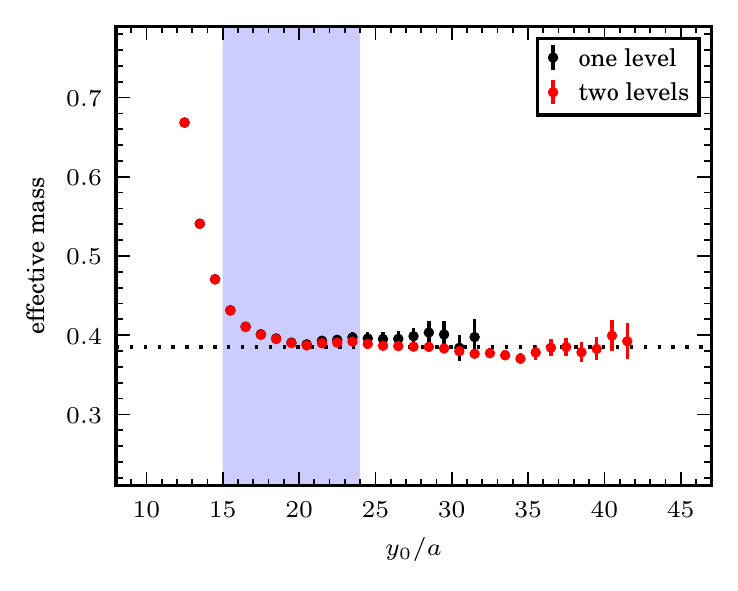}
  \includegraphics[width=0.49\columnwidth]{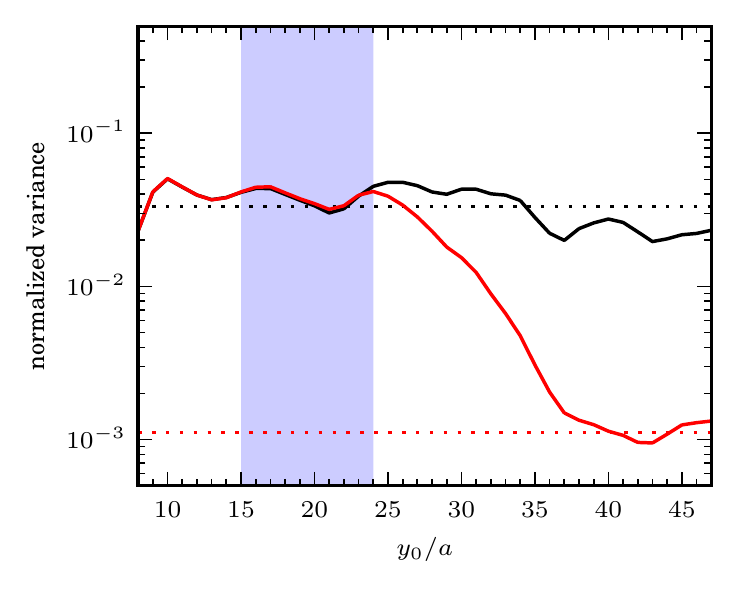}
  \caption{Left: the effective mass of the vector correlator with (red) and
    without two-level integration (black). Right: standard deviation of the vector correlator
    with (red) and without (black) two-level integration both normalized to the standard
    deviation with $n_1=1$.}
  \label{Fig:vector}
\end{figure}

\section{Tests of two-level integration for fermionic correlators}
So far the effectiveness of two-level integration for fermionic correlation
functions has been tested mostly in the quenched approximation of two flavour
QCD. The reason being that the generation of  the gauge field
backgrounds is much cheaper, while keeping the essence of the signal-to-noise ratio
problem. Two-level integration has been implemented for zero momentum correlators of two
singlet pseudoscalar densities~\cite{Ce:2016idq,Ce:2016ajy}, two non-singlet vector
currents~\cite{Ce:lattice2017}, a baryon propagator~\cite{Ce:2016idq},
and meson propagators at non-zero momentum~\cite{Ce:lattice2017}. In all these
exploratory studies an impressive gain in the statistical precision has been observed
when a two-level integration is at work.

Due to lack of space, in this section we will briefly summarize the main results for the
correlators of two non-singlet vector currents and for the baryon propagator only. 
Those correlators have been computed by discretizing gluons and fermions with the Wilson
action, and by imposing open and periodic boundary conditions in the time and spatial
directions respectively~\cite{Luscher:2011kk,Luscher:2012av}. The inverse coupling
constant is fixed to $\beta=6/g_0^2=6.0$, the length of each spatial direction
to $L=24\, a$, and the time extent to $T=64\, a$. The lattice spacing is $a=0.093$~fm as 
fixed by assuming a physical value of $0.5$~fm for the Sommer scale
$r_0/a=5.368$~\cite{Guagnelli:1998ud}. The up and down quarks are taken to be
degenerate with a mass fixed by the hopping parameter value $k=0.1560$,
corresponding to a pion of approximatively $455$~MeV~\cite{Allton:1996yv}.
A $1000$ level-$0$ independent gauge-field configurations have been generated
with the HMC, and for some of them level-1 configurations have been
produced subsequently, see below and Ref.~\cite{Ce:2016idq} for more details.

\subsection{Non-singlet vector two-point function}
For $n_0=50$ of the level-$0$ configurations, $n_1=30$ level-1 gauge fields
have been generated by  updating independently the gauge
field in $\Lambda_0$ and $\Lambda_2$ while freezing the links
in $\Lambda_1$. The latter includes the time slices between $16$ and $23$,
corresponding to a thickness of $\Delta\approx0.7$\,fm and
$M_\pi\Delta \approx 1.7$. On all those configurations the exact Wick contraction
of the non-singlet vector-vector correlator has been computed, for more
details see Ref.~\cite{Ce:lattice2017}.

On the left plot in Fig.~\ref{Fig:vector}, it is shown the effective mass of
the vector correlator as a function of the sink coordinate $y_0$ (the source is
kept fixed at $x_0=8a$) with (red) and without two-level integration (black).
The data are cut when the relative error reaches $10\%$. On the right
plot it is shown the standard deviation of the correlator with (red) and
without (black) two-level integration both normalized to the standard
deviation with $n_1=1$.

A picture similar to the one for the gluonic observables emerges. When the
source and the sink are deep in the two active regions, i.e. $y_0> 24a$,
the square root of the variance is reduced by $\approx 1/n_1$ signaling that
the two-level Monte Carlo is working at full potentiality. With two-level
integration, the plateau in the effective mass turns out to be
approximatively $1$~fm longer than the one computed in the standard way.
No attempt was made to reach the value of $n_1$ at which the reduction
of the variance starts to slow down.

These results suggest that, when applied to full QCD with light quark masses,
the two-level integration can indeed solve the problem of large statistical errors
in the lattice determination of the hadron contributions to the muon $g-2$.

\begin{figure}[t!]
  \centering
  \includegraphics[width=0.95\columnwidth]{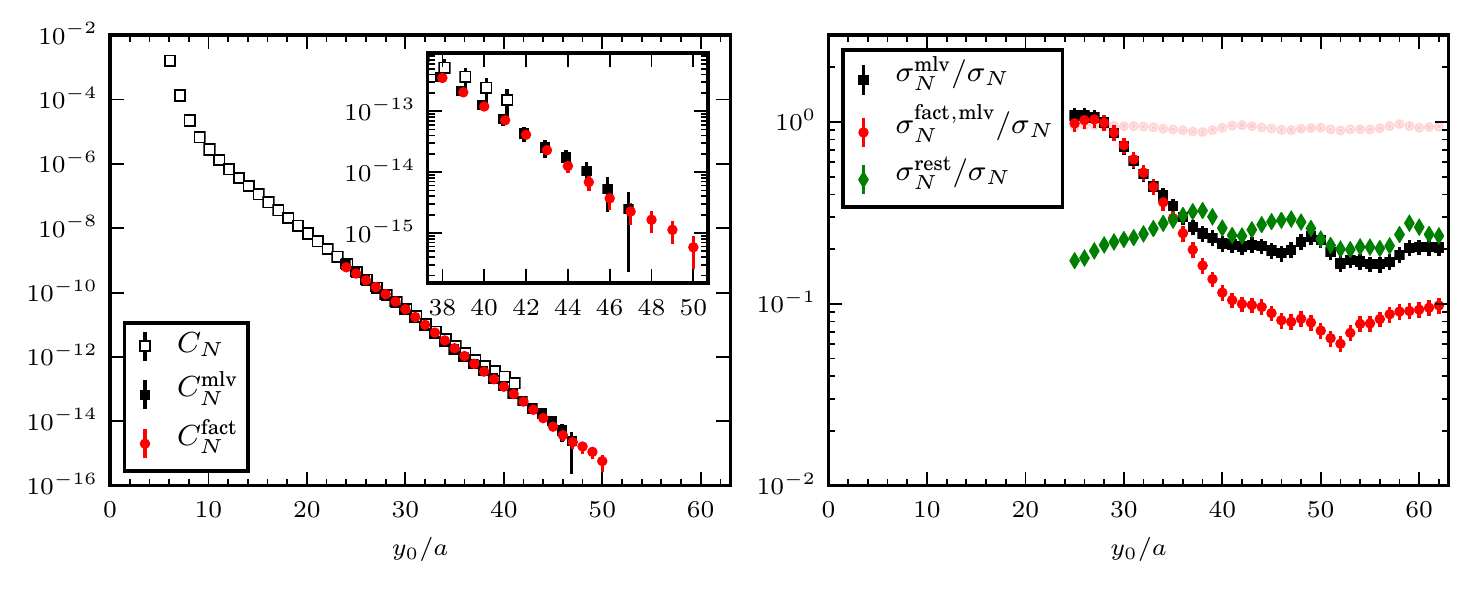}
  \caption{Left: best results for $C_N(y_0,x_0)$ with (filled black squares) and
    without two-level (open squares) integration,
    and for $C^{\,\text{fact}}_N(y_0,x_0)$ only (red circles). Right: standard deviation of
    $C^{\,\text{fact}}_N(y_0,x_0)$ from one-level (light red) and two-level Monte Carlo (red),
    of $C^{\,\text{rest}}_N(y_0,x_0)$ (green),
    and of our best two-level estimate of the exact correlator (black) all normalized to 
    the standard deviation of the one-level estimate of $C_N(y_0,x_0)$.}
    \label{Fig:baryon}
\end{figure}

\subsection{Baryon propagator}
For $n_0=50$ level-$0$ configurations, $n_1=20$ level-1 gauge fields
have been generated by freezing the links in $\Lambda_1$ which, in this case,
includes the time slices between $16$ and $32$, corresponding to a thickness
of\footnote{The larger $\Delta$ chosen here is due to the particular
factorization of the Wick contractions adopted in this case. }
$\Delta\approx 1.4$\,fm and $M_\pi\Delta \approx 3.4$. The gauge fields in 
$\Lambda_0$ and $\Lambda_2$ have then been updated independently.
On all those configurations the exact Wick contractions for the baryon propagator
$C_{N}$, a factorized approximation $C^{\,\text{fact}}_{N}$, and
the remainder defined configuration by configuration by
\be\label{eq:seriesN}
C_{N} = C^{\,\text{fact}}_{N} +
C^{\,\text{rest}}_{N}\; ,   
\ee
have been computed, see Ref.~\cite{Ce:2016idq} for more details. All of them have
been determined starting from local sources on the time-slice at $x_0=4 a$. Extensive
numerical tests show that the factorized correlator approximates the exact one
at the level of $5-10\%$.

The final results for the correlator with (filled black squares) and without
(open squares) the two-level integration are shown
in the left plot of Fig.~\ref{Fig:baryon}, together with the factorized
contribution only (red circles). Thanks to the two-level Monte Carlo, 
the signal-to-noise ratio for the factorized contribution remains larger than
1 for 10 additional time-slices with respect to the standard evaluation. When
the remainder, $C^{\, \text{rest}}_{N}$, is added the gain reduces to 5 additional points.
The effectiveness of the two-level integration is better seen on the right plot
of Fig.~\ref{Fig:baryon}, where the standard deviations of the various contributions
are normalized to the one of the exact correlator. For completeness we report also the
normalized standard deviation on our best two-level estimate of the full correlator.

At large time distances,
the statistical error on the standard estimate of $C_N$ is dominated
by the one on $C^{\, \text{fact}}_N$. Once the two-level integration is switched on, the error on
$C^{\,\text{fact}}_N$ decreases\footnote{With the particular factorization chosen for the baryon propagator,
the origin of the gain is due to various factors, see Ref.~\cite{Ce:2016idq} for a detailed discussion.}
roughly as $n_1^{-1}$, while the one on the remainder continues to scale as $n_1^{-1/2}$. The
multi-level therefore works at its best until the red curve on the right plot of
Fig.~\ref{Fig:baryon} hits the green curve. After that point the statistical error on
the two-level estimate of the correlator is dominated by the one on the remainder, and
increasing $n_1$ is not profitable anymore.

Multi-level simulations of baryon correlation functions solve the problem of the
exponential degradation of the signal-to-noise ratio, and open new perspective
for the computation of baryon masses and matrix elements in lattice QCD.

\section{Conclusions}
The decomposition of the lattice in overlapping domains leads to
a factorization of the gauge-field dependence of the fermion determinant
in QCD. Thanks to a multi-boson representation of the (small) interaction
among gauge fields on distant blocks, the resulting action is local in the
block scalar and gauge fields. It can be efficiently simulated by variants
of the standard hybrid Monte Carlo algorithm. Being the multi-boson contribution
manifestly positive for each single flavour, no additional impediment is
encountered in simulating the strange or heavier quarks. The measurements of local gluonic
observables, such as the energy and the topological charge densities, reveal a
good efficiency of the algorithm in updating the gauge field. No particular
freezing of the links is observed. When combined with the factorization of the
fermion propagator, these results pave the way for multi-level Monte Carlo integration
in the presence of fermions, opening new perspectives in lattice gauge theory.

The numerical tests on gluonic and fermionic correlation functions carried out so
far prove that the signal-to-noise ratio in those computations increases exponentially
with the time distance of the sources when a two-level integration is at work
instead of the standard one-level Monte Carlo. This represents a
turning point for the computation of many interesting quantities sensitive to Standard
Model and hopefully to beyond Standard Model physics: baryon masses and matrix elements
($g_A$, \dots, $<x>_{u-d}$), the hadronic contributions to the muon $g-2$, leptonic and
semi-leptonic $B$ decays, $\rho$ , $\eta'$, etc.

The factorization does not require a particular shape of the domains, nor does each
of them need to be connected. What matters is a minimum distance of $\approx 0.5$~fm
among the blocks which are active during the level-1 updates. It is therefore already quite
clear that its generalization to four dimensions would localize the simulations of theories with fermions,
allow for very large volumes to be generated in master-field simulations~\cite{Luscher:2017cjh},
and open lattice QCD to a new class of physics problems.

The  proposed method  relies on two key ingredients: the locality of the Wilson Dirac operator and
a (configuration by configuration) fast decrease of its inverse with the distance between
the sink and the source. The ideas and the computational strategy presented here may, therefore,
be applicable to very different theories with fermions if they enjoy these very basic properties. 

\section{Acknowledgments}
L.~G. thanks M. L\"uscher for inspiring discussions at CERN during the
preparation of this talk, and for sharing the unpublished notes on the relation
between the determinant factorization in Eq.~(\ref{eq:factfinal}) and the
one obtained when an overlapping Schwarz preconditioner is used for the Dirac
operator~\cite{Luscher:fact}. Many thanks to O. B\"ar for sharing some of
the information of his plenary talk before the conference. L.~G. thanks
C. Lehner and M. Bruno for an interesting discussion on the hadron vacuum polarization contribution
to the muon $g-2$. Many thanks to the organizers of the Lattice 2017 conference
for preparing an outstanding scientific program in the beautiful Granada.

\bibliography{mb}

\end{document}